\documentclass[twocolumn,showpacs,preprintnumbers,amsmath,amssymb]{revtex4}

\usepackage{graphicx}
\usepackage{dcolumn}
\usepackage{bm}

\begin{document}
\newlength{\figwidth}
\setlength{\figwidth}{0.5 \textwidth}
\addtolength{\figwidth}{-0.5 \columnsep}
\addtolength{\figwidth}{-1cm}

\preprint{}

\title{Methane Hydrate under High Pressure: \\
Searching for Centering of Hydrogen Bonds
}

\author{Toshiaki Iitaka}
 \email{tiitaka@riken.jp}  
 \homepage{http://atlas.riken.go.jp/~iitaka}
\author{Toshikazu Ebisuzaki}%
\affiliation{
Computational Science Division, \\
RIKEN (The Institute of Physical and Chemical Research) \\
2-1 Hirosawa, Wako, Saitama 351-0198, Japan} 

\date{\today}

\begin{abstract}
The structural, electronic, and spectroscopic properties of a high-pressure phase of methane hydrate (MH-III) are studied by first principles electronic structure calculations.  A detailed analysis of the atomic positions suggests that {\it ionization} of hydrogen-bonded water molecules occurs around 40GPa and {\it centering} or symmetrization of hydrogen-bonds occurs around 70 GPa. These pressures are much lower compared with ioninzation around 55 GPa and centering around 100 GPa in pure ice. The transition may be observed with low-temperature IR/Raman spectroscopy of OH stretching modes or neutron diffraction. 
\end{abstract}

\maketitle


Methane hydrate (MH), known as \textit{Burning Ice}, is a special class of ice that contains methane molecules in cages or networks of hydrogen bonded water molecules.
Low pressure phase of methane hydrate (MH-I) forms sI structure of cages\cite{Sloan1998}. MH-I, abundant in the deep ocean, has been attracting attention of the industry as a key material of new energy resource, whose amount is estimated twice as much as the total fossil fuel reserve\cite{Sloan1998}.
 
MH is also known as an important material for understanding the mystery of the atmosphere of Titan, the largest satellite of Saturn. The conventional theory \cite{Lunine1985} could not explain abundant methane gas in Titan's atmosphere because MH-I inside Titan was assumed to decompose into ice and methane around 1 or 2 GPa, and escape to the atmosphere to be photodecomposed in early stage of Titan's history. To understand this mystery is one of the goals of the Cassini-Huygens spacecraft, which started its journey in 1997 and will arrive at Saturn system in 2004 \cite{JPL}.

On the earth, in 2001, Loveday \textit{et al.} \cite{Loveday2001,Loveday2001b} discovered new phases of MH by X-ray and neutron diffraction experiments under high pressure: MH-I transforms to MH-II ( sH cage structure) at 1 GPa, and then to MH-III phase (orthorhombic filled ice structure) at 2 GPa, which survives at least up to 10 GPa. Other researchers reported similar high pressure phases\cite{Hirai2001,Chou2001}. Recently Hirai \textit{et al.} \cite{Hirai2003} reported that MH-III survives up to 42 GPa at room temperature. Shimizu \textit{et al.} \cite{Shimizu2002} have measured the site- and pressure-dependence of CH- and OH-vibration frequencies in these phases up to 5.2 GPa. 
Discovery of these high pressure phases allows us a new explanation of abundant methane gas in Titan's atmosphere: methane gas may be reserved in thick layers of MH-III under Titan's surface and gradually emit to the atmosphere from the reservoir\cite{Loveday2001}.

\begin{figure*}[hbt]
\resizebox{\textwidth}{!}{
\includegraphics{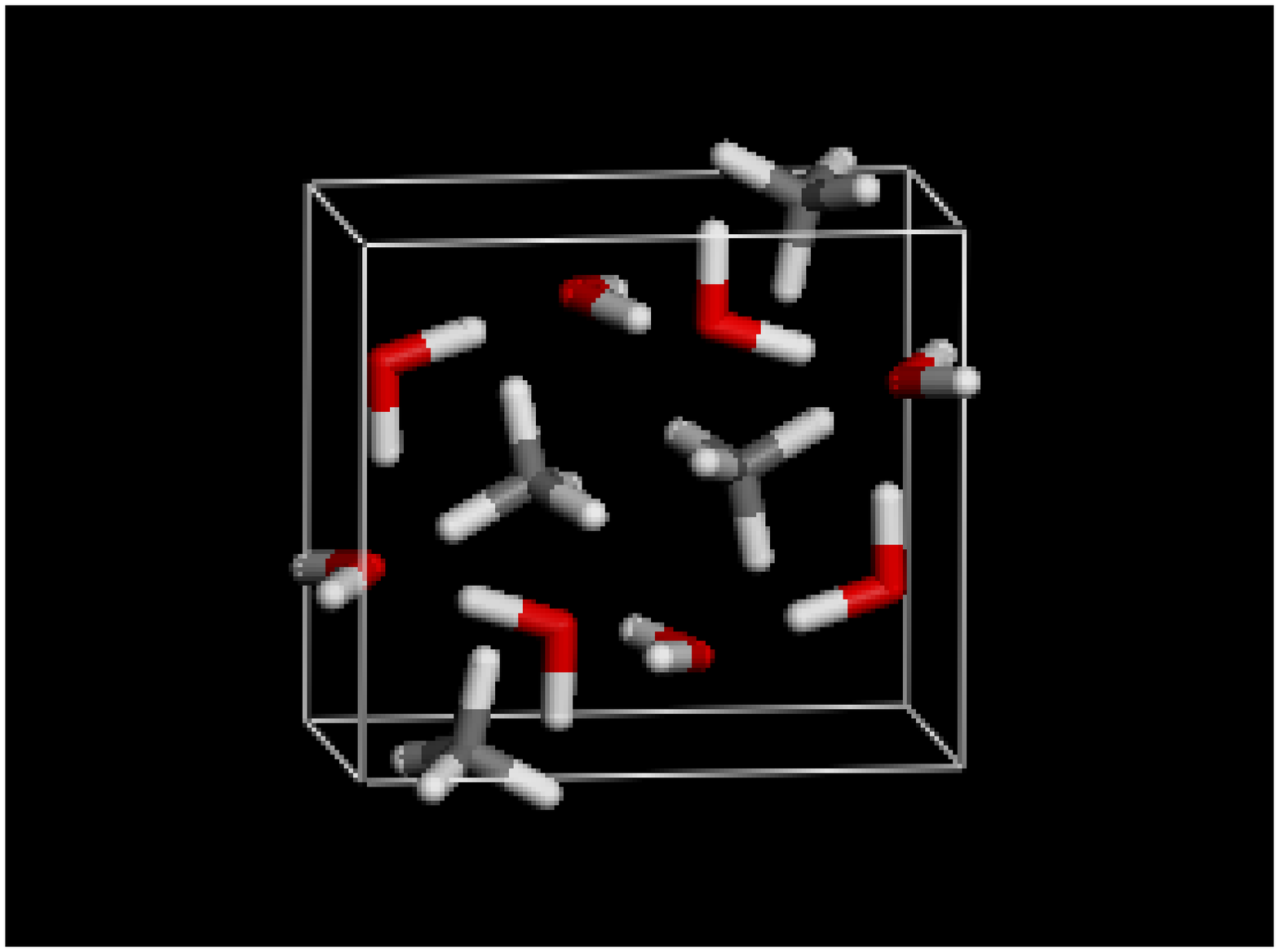}
\includegraphics{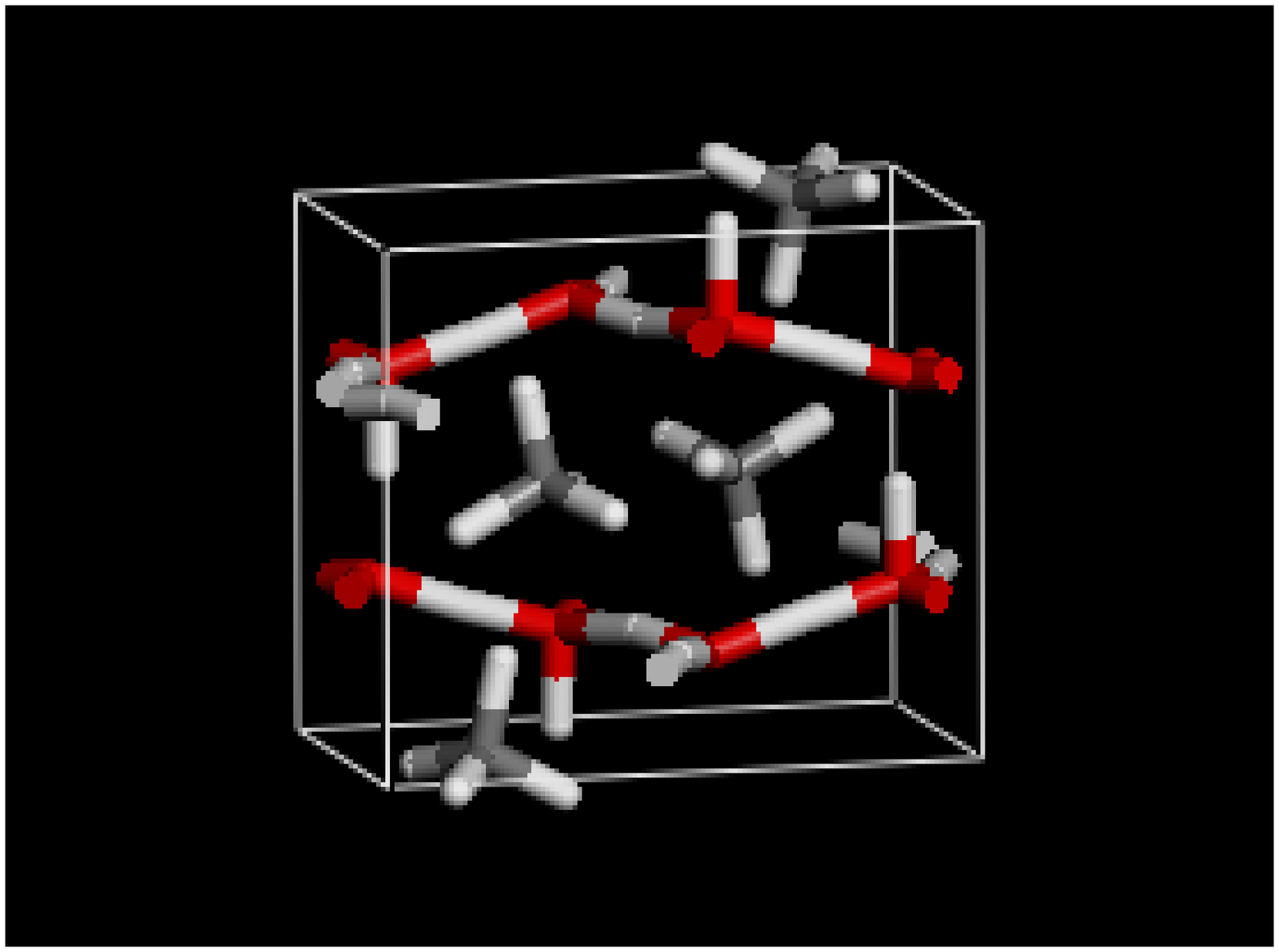}
\includegraphics{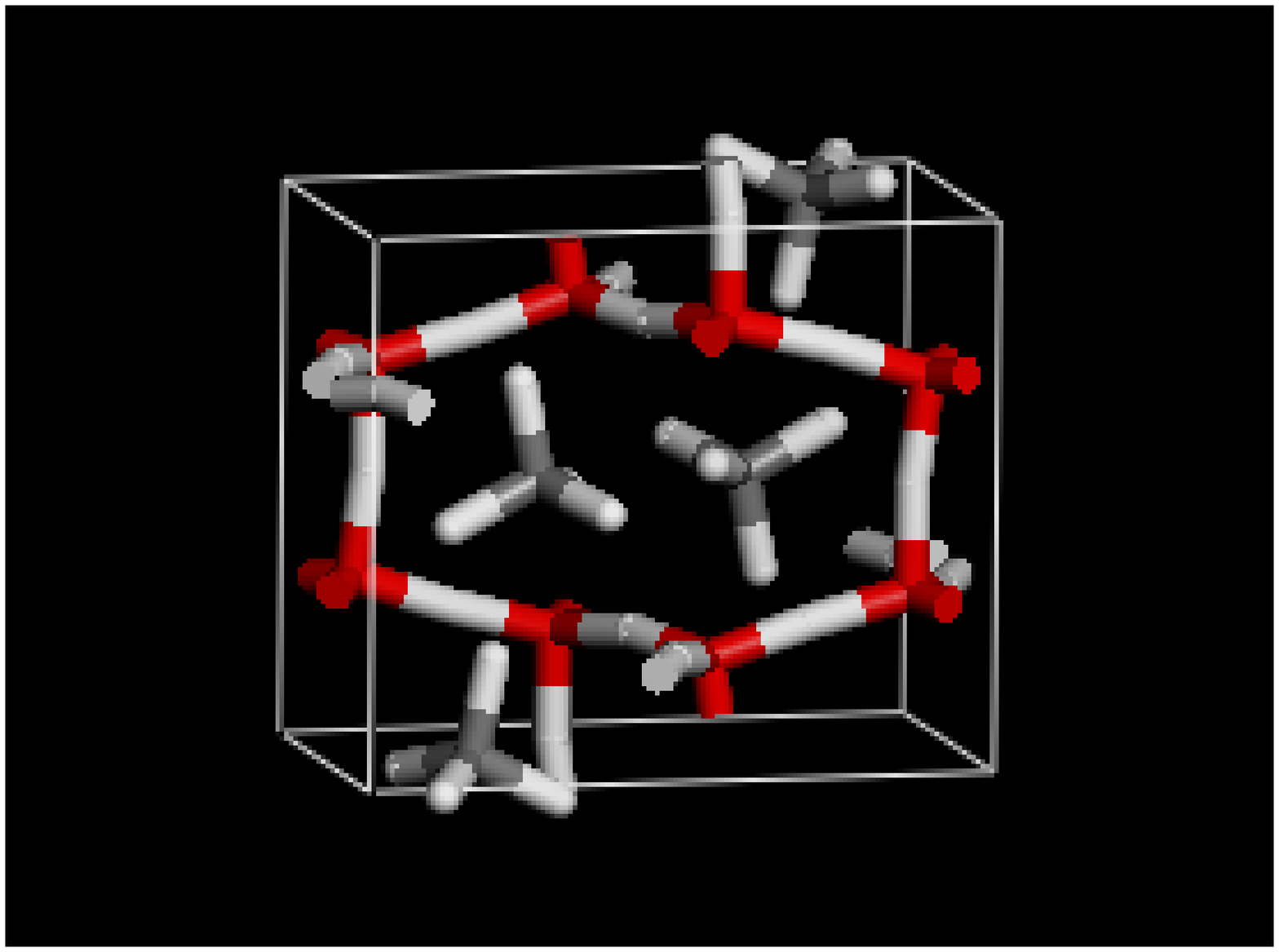}
}
\caption{Crystal Structure of MH-III (a) 40 GPa, (b) 60 GPa, (c) 80 GPa}
\label{fig:structure40}
\end{figure*}

In this article, we focus on the features of MH-III as a promising material for investigateing {\it centering} or symmetrization of hydrogen bonds between water molecules. Advantage of using MH-III for studying the centering is that it is expected to occur at much lower pressure than that of pure ice (ice VII-ice X transition)\cite{Petrenko1999,Jeffrey1997,Benoit2002,Loubeyre1999,Goncharov1999,Struzhkin1997,Aoki1996a,Hirsch1986,Benoit1998,Bernasconi1998,Putrino2002}, which may make the difficult high-pressure experiments easier. 
In the following, we calculate the crystal structure and  vibrational spectra of MH-III by using the density functional theory, so that we can predict and analyze experimental results.

We modeled the crystal structure of MH-III at 3 GPa by using the $Pmcn$ symmetry, the lattice parameters and the position of atoms determined by the diffraction experiment \cite{Loveday2001b}. We chose the orientation of methane molecules carefully so that the molecules become close packed.  The model consists of four methane and eight water molecules in the unit cell, of which one methane and two water molecules are symmetrically inequivalent. 
This structure was found stable after full geometrical optimization, in which the enthalpy $H=E+PV$ was minimized by varying the lattice vectors and the positions of atoms without any constraints such as crystal symmetry. The structures at higher pressures ( Figs.~\ref{fig:structure40} ) were calculated in a similar manner.

\begin{figure}
\resizebox{\figwidth}{!}{\includegraphics{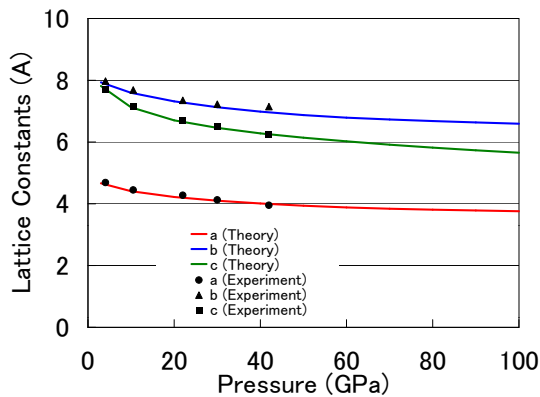}}
\caption{Lattice parameters: The solid lines indicate the present density functional theory calculation. The symbols indicate the experimental results of Hirai {\it et al.} \cite{Hirai2003} }
\label{fig:lattice}
\end{figure}

The details of the electronic structure calculation \cite{Payne1992} are as follows: the valence wave functions are expanded in a plane wave basis set truncated at a kinetic energy of 1520 eV. The electron-ion interactions are described by the Vanderbilt-type ultrasoft pseudopotentials \cite{Vanderbilt1990}. The effects of exchange-correlation interaction are treated within the generalized gradient approximation of Perdew \textit{et al.} (GGA-PBE) \cite{Perdew1996}. The Brillouin zones are sampled with $4 \times 2 \times 2$ Monkhorst-Pack k-points \cite{Monkhorst1976} by using time-reversal symmetry only. In the geometrical optimization, the total stress tensor is reduced to the order of 0.01 GPa by using the finite basis-set corrections \cite{Francis1990}.

Figure~\ref{fig:lattice} shows the pressure dependence of the calculated lattice parameters at the zero temperature. It agrees well with the experimental data at room temperature by Hirai et al. \cite{Hirai2003}. The cell is the most soft along the c-axis. This feature is also evident from the Figures 1. The compression of the cell along the c-axis is mainly caused by the flattening of the two graphite-like wrinkled sheets normal to c-axis. As a result, the hydrogen bond network at very high pressures becomes $sp_2$-like structure in contrast to $sp_3$ structure of ice Ih.

\begin{figure}
\resizebox{\figwidth}{!}{\includegraphics{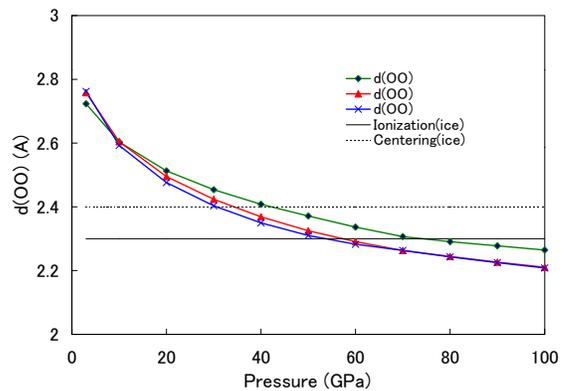}}
\caption{Three symmetrically different distances between oxygen atoms, $d(OO)$, as a function pressure: the dashed and solid horizontal lines indicate the $d(OO)$ at which ionization and centering occur in pure ice, respectively \cite{Benoit2002}.}
\label{fig:OO}
\end{figure}

\begin{figure}
\resizebox{\figwidth}{!}{\includegraphics{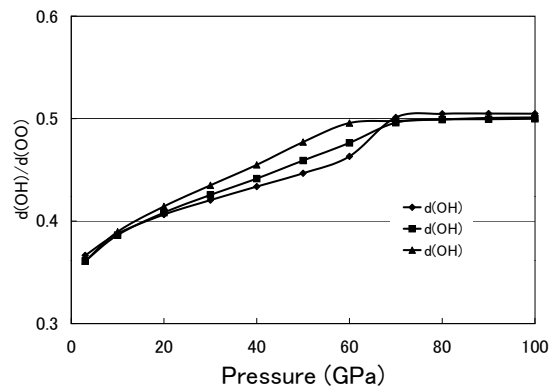}}
\caption{Hydrogen position: The ratio $d(OH)/d(OO)$  calculated by density functional theory is shown as a function of pressure. }
\label{fig:OH}
\end{figure}

In pure ice, centering of the hydrogen bond $O-H \cdots O$ occurs as the  oxygen-oxygen distance $d(OO)$ decreases with pressure increase \cite{Petrenko1999,Jeffrey1997}. At large $d(OO)$  the proton occupies one of two potential minima along the $O-H \cdots O$ bond. As $d(OO)$ becomes smaller the barrier between the two minima becomes lower. When $d(OO)$ becomes smaller than some critical distance, the proton occupies the single minimum at the midpoint between the two oxygen atoms (centering). In the following, we neglect the effect of finite temperature and quantum nature of hydrogen atom unless otherwise stated.
Figure~\ref{fig:OO} shows $d(OO)$  as a function of pressure. There are three curves corresponding to the three symmetrically inequivalent $O-H \cdots O$ bonds. Around 70 GPa $d(OO)$'s become sufficiently short for the hydrogen bond centering in pure ice \cite{Benoit2002}. Indeed Figures~\ref{fig:structure40} show that the centering occurs around this pressure.  Figure~\ref{fig:OH} shows the ratio $d(OH)/d(OO)$  as a function of pressure. The ratio starts from 0.35 (water molecule) at 3 GPa and reaches to 0.5 (centering) around 70 GPa. 

\begin{figure}
\resizebox{\figwidth}{!}{\includegraphics{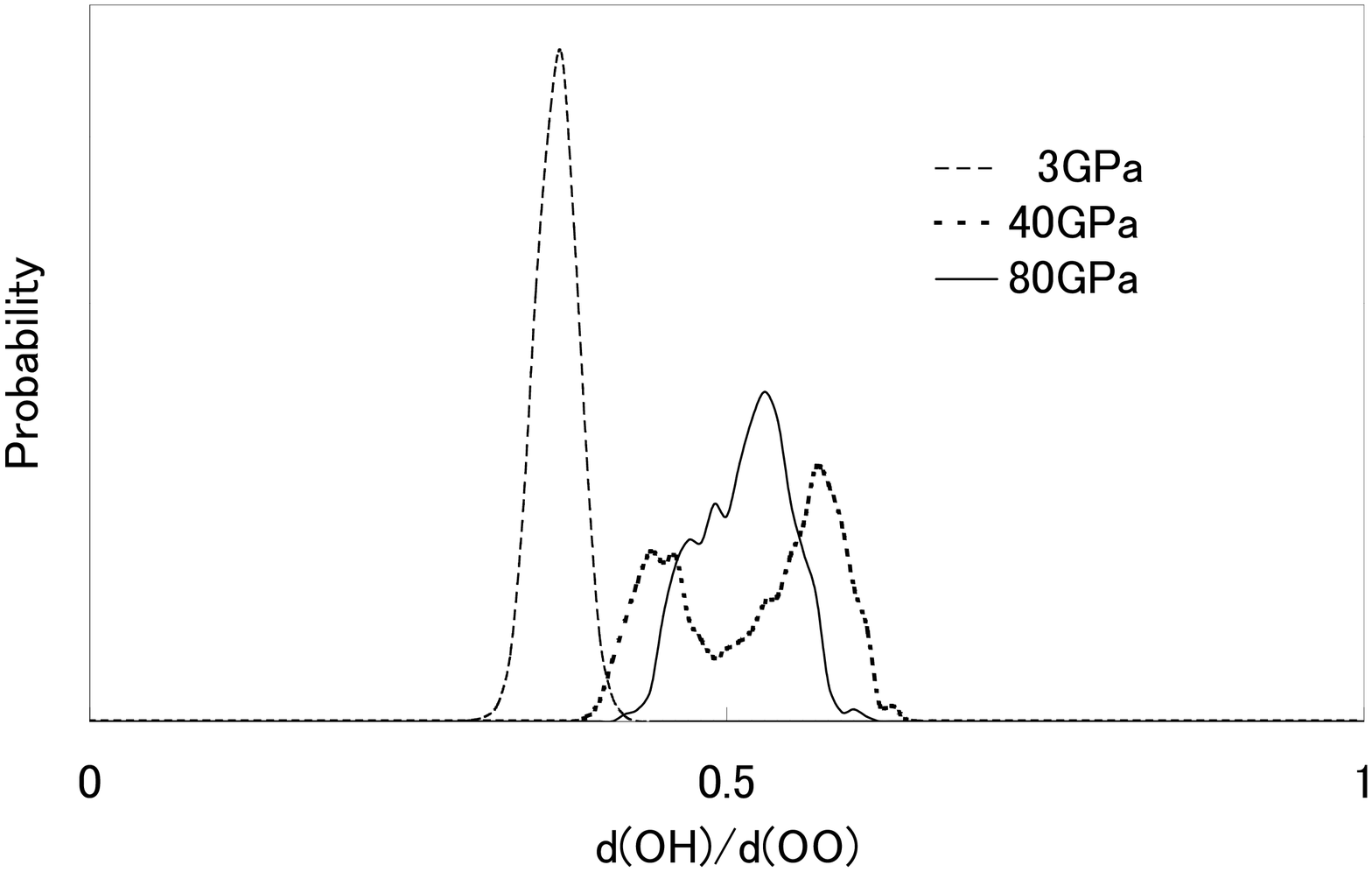}}
\caption{Distribution of hydrogen atom in a hydrogen bond of MH-III at 300K.}
\label{fig:delta}
\end{figure}

Now, let us look into the thermal effects. Benoit {\it et al.} \cite{Benoit2002} recently proposed {\it three-stage scenario} for explaining the hydrogen bond centering \cite{Loubeyre1999} of pure ice with increase of pressure at room temperature: ice stays in {\it molecular state} under low pressures where $H_2O$ remain water molecule, then hydrogen atoms start to jump between two potential minima in {\it ionized state} under midium pressures, finally hydrogen atoms sit at the midpoint between the oxygen atoms in {\it centering state} under high pressures. The $d(OO)$ at which ionization and centering occur in pure ice are indicated dashed and solid horizontal lines in Fig.~\ref{fig:OO}. From this figure we can read that the ionization and cengering in MH-III are expected around 40 GPa and 70 GPa, respectively. The probability distributions of $d(OH)/d(OO)$ for a hydrogen bond of MH-III at room temperature and with pressure 3 GPa, 40 GPa, and 80 GPa are calculated with Car-Parrinello molecular dynamics \cite{cpmd} (Fig.~\ref{fig:delta}), which clearly shows that {\it three stage senario} is also valid in MH-III but with much lower pressure than pure ice.

\begin{figure}
\resizebox{\figwidth}{!}{\includegraphics{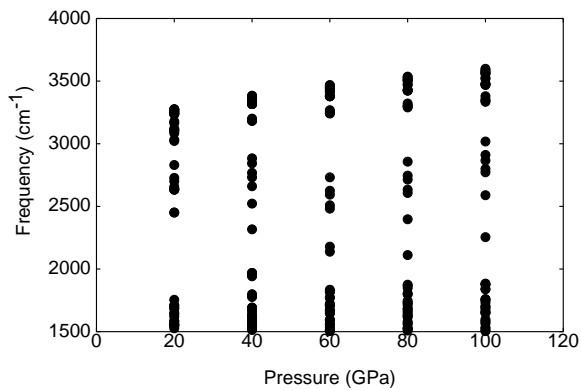}}
\caption{Normal mode frequencies calculated with the density functional linear response theory as a function of pressure.}
\label{fig:normalmode}
\end{figure}

Figure~\ref{fig:normalmode} shows the pressure dependence of normal mode frequencies of MH-III calculated with the density functional linear response theory \cite{Gonze1997}. The modes between 3000 ${\rm cm}^{-1}$ and 3500 ${\rm cm}^{-1}$ are the CH vibrations, whose frequencies monotonically increase with pressure. The modes between 2000 ${\rm cm}^{-1}$ and 3000 ${\rm cm}^{-1}$ are OH-stretching modes. Their frequencies decrease monotonically up to 70 GPa, where centering occurs, and then start to increase. This tendency is qualitatively in accordance with that of pure ice \cite{Goncharov1999}.

\begin{figure}

\resizebox{\figwidth}{!}{
\includegraphics{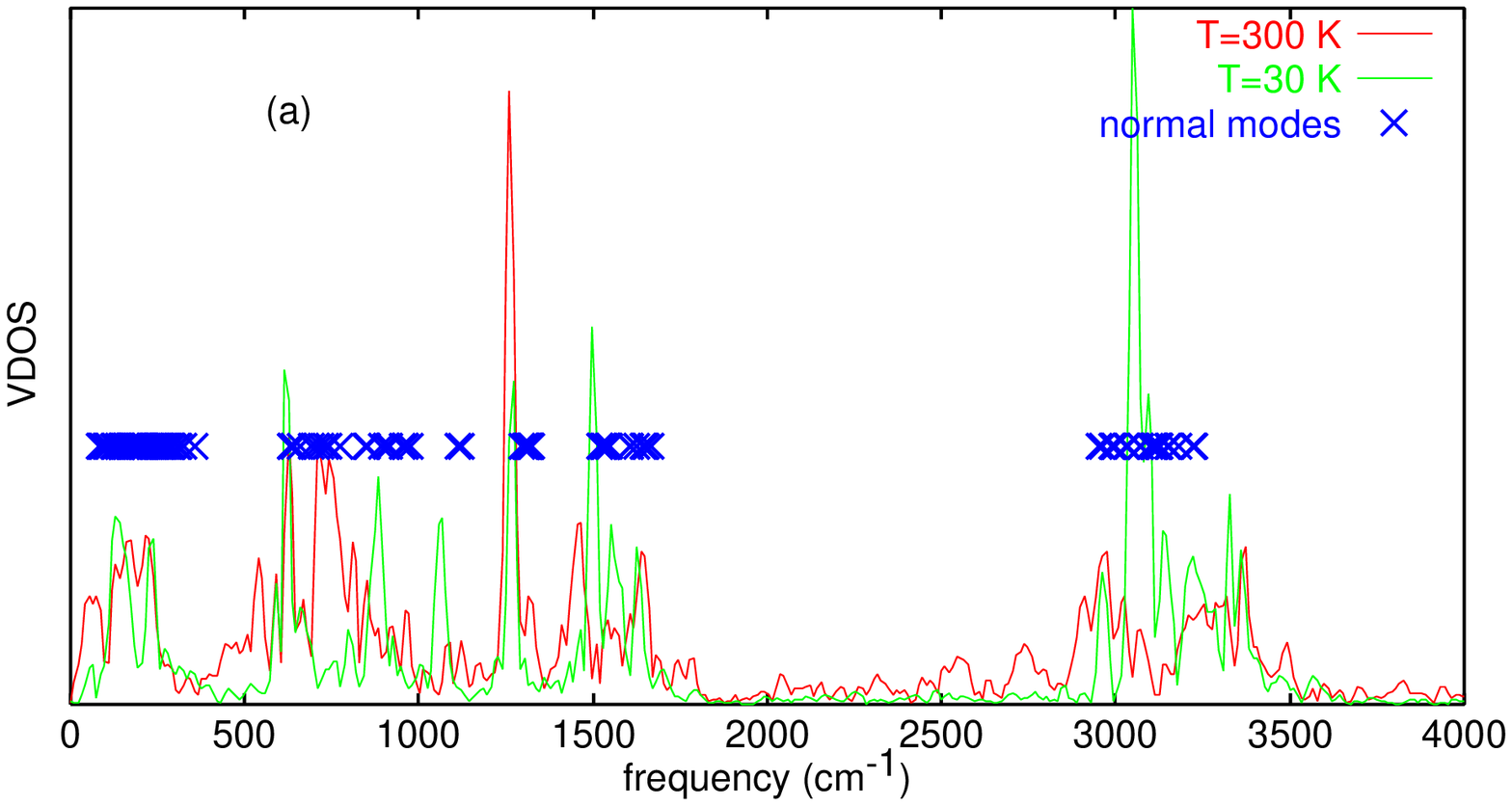}
}

\resizebox{\figwidth}{!}{
\includegraphics{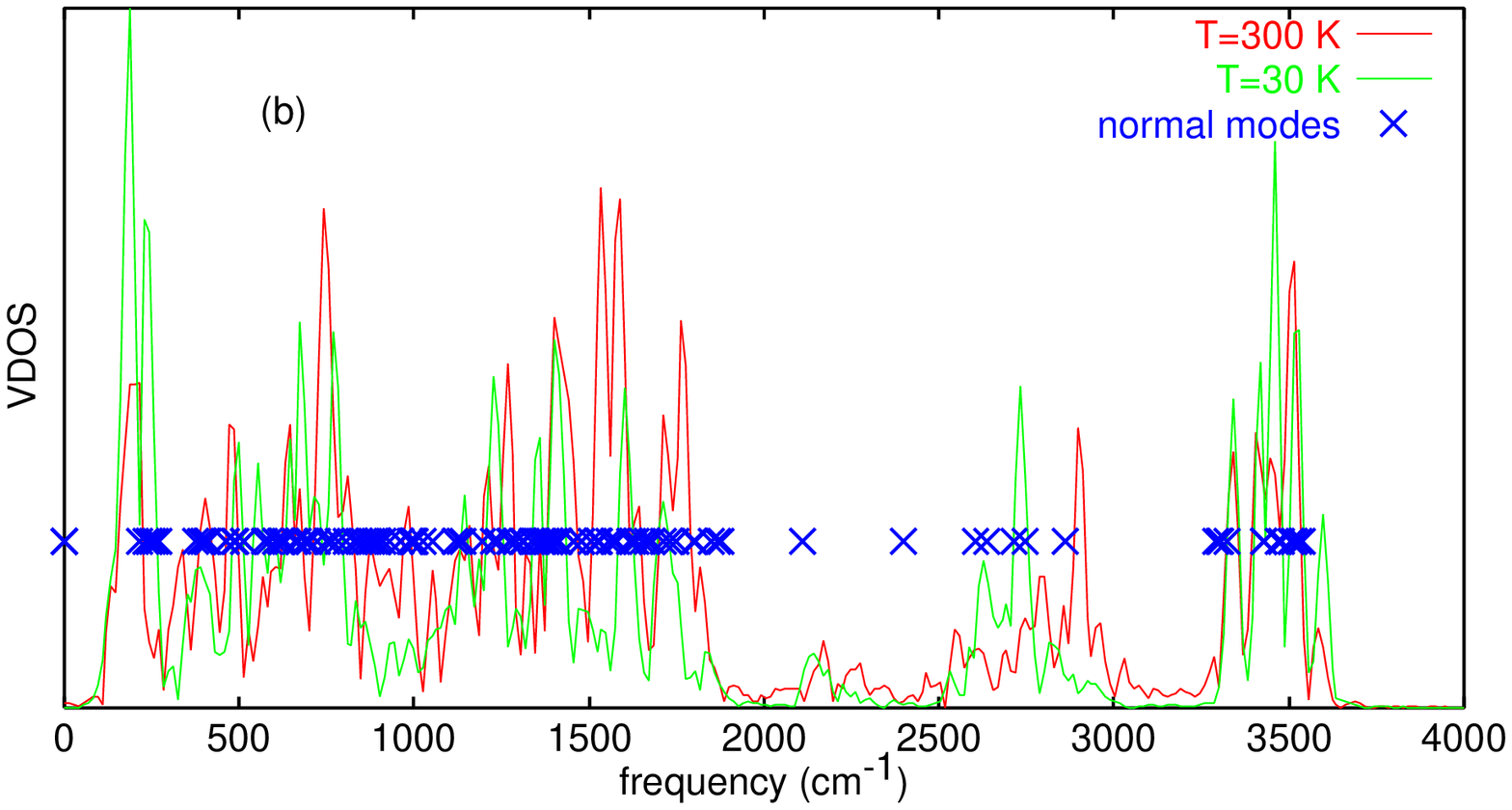}
}
\caption{Vibrational spectra of MH-III at (a) 3 GPa and (b) 80 GPa calculated by  Car-Parrinello Molecular Dynamics are compared with the normal mode frequencies calculated with the density functional linear response theory. Solid line indicates result of CPMD at 300K, dashed line at 30K, and crosses are normal mode frequencies calculated with the density functional linear response theory\cite{Gonze1997}.}
\label{fig:spectrum}
\end{figure}

In Figure~\ref{fig:spectrum}(a),  vibrational spectra of MH-III at 3 GPa are calculated as a Fourier transform of velocity-velocity correlation function obtained from  Car-Parrinello molecular dynamics \cite{cpmd} and compared with the normal mode frequencies calculated with the density functional linear response theory\cite{Gonze1997}.. The peak around 3100 ${\rm cm}^{-1}$ characteristic to OH-stretching vibration of hydrogen bonded water molecules is prominent at 30 K but disappears at 300K.  This fact is consistent with the experimental observation that the Raman peak around 3100 ${\rm cm}^{-1}$ disappeared in MH-III at room temperature \cite{Shimizu2002}, and can be interpreted as the result of strongly anharmonic potential. 
In Figure~\ref{fig:spectrum}(b) shows  vibrational spectra of MH-III at 80 GPa. The OH-stretching modes of centered hydrogen bonds are located between 2000 ${\rm cm}^{-1}$ and 3000 ${\rm cm}^{-1}$. At low temperature the peaks agrees well with the normal modes, while the peaks are blue-shifted at room temperature probably due to weak anharmonicity. 
The dissapearance of the peak of the OH-stretching mode in molecular and ionized states at room temperature and the survival of the peak due to OH-stretching mode in centered hydrogen bond in MH-III are analogous to those in ice VII and ice X \cite{Aoki1996a}.

\begin{figure}
\resizebox{\figwidth}{!}{\includegraphics{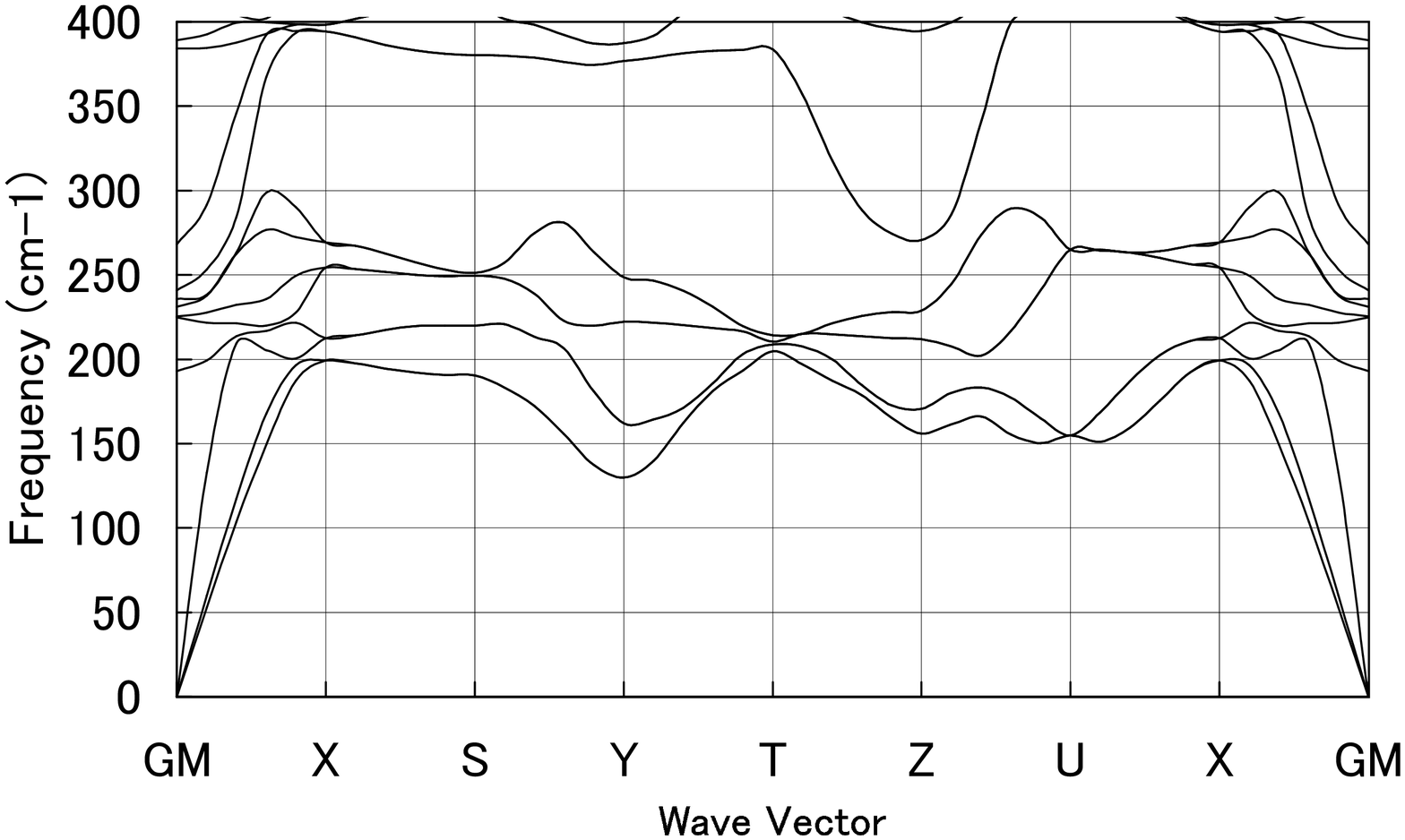}}
\caption{
Phonon dispersion of MH-III at 80 GPa calculated with the density functional linear response theory\cite{Gonze1997}.
}
\label{fig:phonon}
\end{figure}

Figure~\ref{fig:phonon} shows the phonon dispersion of MH-III at 80 GPa calculated with the density functional linear response theory \cite{Gonze1997}.  All frequencies at all wave numbers have positive frequencies indicating the mechanical stability  of this structure at zero temperature. However, further experimental and theoretical studies are necessary to prove the thermal stability at room temperature.


In summary, we studied the structural and spectral properties of high pressure phase of methane hydrate (MH-III) with the density functional theory, and showed that ionization and centering of hydrogen bonds in MH-III may be observed around 40 GPa and 70 GPa, respectively, which are much lower pressure than those of pure water ice. Therefore MH-III may provide precious information about ionization and centering of hydrogen bonds between water molecules as the second example after pure ice. From the view point of planetary science, studying physical and chemical properties of clathrate hydrate such as MH will become more and more important to interpret and to understand the information of the outer solar systems sent from space-missions such as Cassini-Huygens spacecraft arriving at Saturn system in 2004.

\begin{acknowledgments}
One of the authors (TI) would like to thank Dr. Hiroyasu Shimizu, Dr. Shigeo Sasaki, Dr. John S. Tse, Dr. Hisako Hirai, Dr. Takashi Ikeda, and Dr. Akira Hori for invaluable discussions.
The results presented here were computed by using supercomputers at RIKEN and NIG.
\end{acknowledgments}

\bibliography{mh3}

\end{document}